\documentclass[sigconf,natbib=false]{acmart}
\usepackage{booktabs}
\usepackage{multirow}
\usepackage{amsmath}
\usepackage{array}
\AtBeginDocument{%
  }

\setcopyright{acmlicensed}
\copyrightyear{2024}
\acmYear{2024}
\acmDOI{XXXXX.XXXXX}

\acmConference[CIKM'24]{33rd ACM International Conference on
Information and Knowledge Management}{October 21--25,
  2024}{Boise, ID}
\acmISBN{978-1-4503-XXXX-X/18/06}

\RequirePackage[
  datamodel=acmdatamodel,
  style=acmnumeric,
  ]{biblatex}

\begin{document}

\title{UniRec: A Dual Enhancement of Uniformity and Frequency in Sequential Recommendations}

 \author{Yang Liu}
 \email{liuyangfd22@m.fudan.edu.cn}
 \orcid{0009-0004-4380-6888}
 \affiliation{%
   \department{School of Computer Science}
   \institution{Fudan University}
   \city{Shanghai}
   \country{China}}
 \author{Yitong Wang}
 \email{yitongw@fudan.edu.cn}
 \authornotemark[1]
 \affiliation{%
   \department{School of Computer Science}
   \institution{Fudan University}
   \city{Shanghai}
   \country{China}}
 \author{Chenyue Feng}
 \email{cyfeng22@m.fudan.edu.cn}
 \orcid{0009-0007-0018-9164}
 \affiliation{%
   \department{School of Computer Science}
   \institution{Fudan University}
   \city{Shanghai}
   \country{China}}

\renewcommand{\shortauthors}{Liu et al.}
\begin{abstract}
Representation learning in sequential recommendation is critical for accurately modeling user interaction patterns and improving recommendation precision. However, existing approaches predominantly emphasize item-to-item transitions, often neglecting the time intervals between interactions, which are closely related to behavior pattern changes. Additionally, broader interaction attributes, such as item frequency, are frequently overlooked. We found that both sequences with more uniform time intervals and items with higher frequency yield better prediction performance. Conversely, non-uniform sequences exacerbate user interest drift and less-frequent items are difficult to model due to sparse sampling, presenting unique challenges inadequately addressed by current methods. In this paper, we propose UniRec, a novel bidirectional enhancement sequential recommendation method. UniRec leverages sequence uniformity and item frequency to enhance performance, particularly improving the representation of non-uniform sequences and less-frequent items. These two branches mutually reinforce each other, driving comprehensive performance optimization in complex sequential recommendation scenarios. Additionally, we present a multidimensional time module to further enhance adaptability. To the best of our knowledge, UniRec is the first method to utilize the characteristics of uniformity and frequency for feature augmentation. Comparing with eleven advanced models across four datasets, we demonstrate that UniRec outperforms SOTA models significantly. The code is available at https://github.com/Linxi000/UniRec.
\end{abstract}

\ccsdesc[500]{Information systems~Recommender systems}
\keywords{Sequential Recommendation, Sequence Uniformity, Item Frequency, Feature Enhancement}

%
\maketitle

\section{Introduction}
\begin{figure}
    \centering
    \includegraphics[width=\columnwidth]{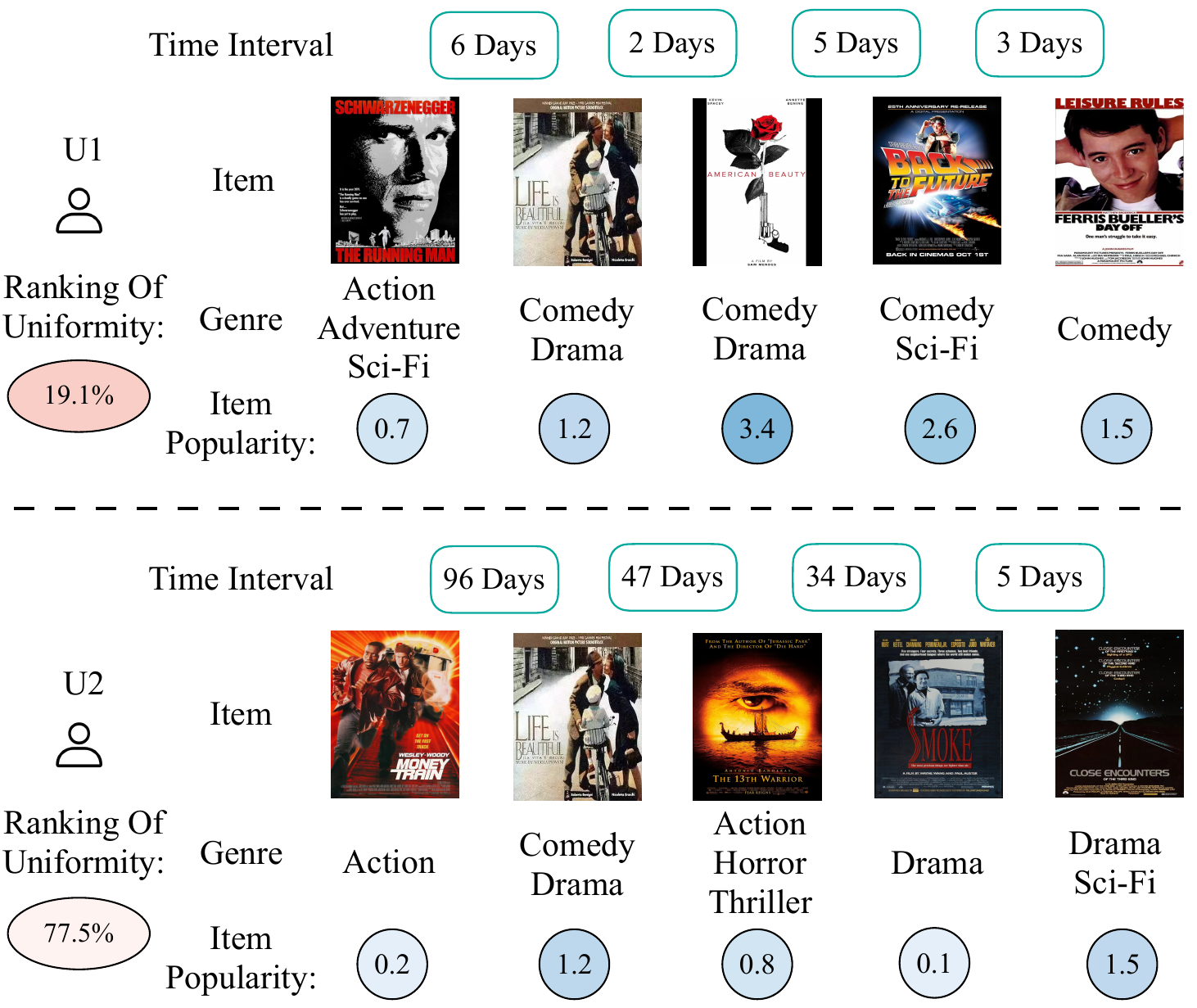}
    \caption{An example of uniform and non-uniform sequences in a real dataset.}
    \label{fig:fig1}
\end{figure}
Sequential recommendation systems have become increasingly prevalent due to their ability to effectively model user preferences \cite{wang2019sequential,quadrana2018sequence,fang2020deep,wang_survey_2022}. Such systems utilize the sequential order of user interactions over time to predict future interests \cite{kang_self-attentive_2018,sun_bert4rec_2019,fan_lighter_2021}. Incorporating temporal information into these algorithms has proven effective, as it provides significant insights into user behavioral patterns \cite{li_time_2020,ye_time_2020,cho_meantime_2020,fan_continuous-time_2021,tran_attention_2023,du_frequency_2023}. Current approaches primarily focus on modeling explicit timestamps \cite{li_time_2020,rahmani_incorporating_2023} or capturing cyclic patterns \cite{cho_meantime_2020}, but they often overlook time intervals, which reveal user characteristics and convey critical information within user interaction sequences. Yizhou Dang et al. propose that variations in the time intervals between sequential interactions can serve as indicators of shifts in user preferences \cite{dang_uniform_2023}. Building on this premise, they designed data augmentation operators to improve the uniformity of sequences. However, this direction still lacks full study and holds potential significance, as sequence uniformity is a common phenomenon across various datasets. Additionally, the effectiveness of a model in capturing item characteristics is influenced by the frequency of these items. While considerable research has focused on enhancing the recommendation performance for long-tail items \cite{kim2019sequential,liu2020long}, the utilization of item frequency to enhance model performance remains an area requiring further exploration.

Figure \ref{fig:fig1} illustrates segments of the interaction of uniform sequence versus non-uniform sequences from different users, encompassing items of both high and low frequency. The "Ranking of Uniformity" sorts interaction sequences by the variance of their time intervals in ascending order, with lower percentages indicating greater uniformity. For example, U1 with a ranking of 19.1\% is more uniform than 80.9\% of the sequences. "Item Popularity" is defined as the proportion of an item's occurrences relative to the number of all interactions, thus quantifying the frequency of item appearances within the dataset. This figure illustrates that time intervals within uniform sequences are typically shorter and more stable, indicating steadier user interests. In contrast, non-uniform sequences exhibit more variable time intervals, reflecting more frequent changes in user interests. Furthermore, the intensity of the color within the circles signifies the model’s effectiveness in learning the representations of the corresponding users or items, with darker colors indicating higher effectiveness.

We first analyze the performance of sequences with different intervals and item frequencies in section \ref{sec:2} and validate that sequences with higher uniformity and items with greater frequency tend to exhibit better performance. Following this, we implement a dual enhancement approach UniRec in section \ref{sec:3}. For sequences, we generate non-uniform subsets from uniform sequences by incorporating less-frequent items to simulate fluctuating user interests, thereby enhancing the modeling of non-uniform sequence representations later. For items, we train a neighbor aggregation mechanism on frequent items and extend it to less-frequent items using curriculum learning to improve their representations and transfer this knowledge to sequence modeling. This dual-branch approach is simple and effective, providing a new perspective for feature enhancement in sequential recommendation. Additionally, we integrate the temporal characteristics of both uniform and non-uniform sequences to conduct multidimensional temporal modeling.

In summary, the contributions of this paper are as follows:

\begin{itemize}
\item We propose a novel dual enhancement architecture that leverages sequence uniformity and item frequency. This architecture comprises two independent yet mutually reinforced branches, collectively driving comprehensive performance optimization.
\item We improve the model's ability to handle non-uniform sequences and less-frequent items and provide a new perspective for feature enhancement in sequential recommendation.
\item We conduct extensive experiments on 4 real-world datasets, demonstrating significant improvements over 11 competing models, including 6 cutting-edge models that incorporate temporal modeling in their sequential recommendation systems.
\end{itemize}

\section{Preliminary Study} \label{sec:2}
\begin{table}[htbp]
\setlength{\tabcolsep}{1pt}

\scriptsize
\centering
\caption{Performance of sequential recommendation models on different subsets.}
\label{table:1}
\setlength{\tabcolsep}{1.15 pt} 
\begin{tabular}{@{}lcccccccccccc@{}}
\toprule
\multirow{2}{*}{\textbf{Dataset}} & \multirow{2}{*}{\textbf{Strategy}} & \multicolumn{3}{c}{\textbf{SASRec}} & \multicolumn{3}{c}{\textbf{Bert4Rec}} & \multicolumn{3}{c}{\textbf{LightSANs}} \\ 
\cmidrule(r){3-5} \cmidrule(r){6-8} \cmidrule(r){9-11}
& & \textbf{NDCG} & \textbf{Hit} & \textbf{MRR} & \textbf{NDCG} & \textbf{Hit} & \textbf{MRR} & \textbf{NDCG} & \textbf{Hit} & \textbf{MRR} \\ 
\midrule
\multirow{7}{*}{\textbf{ML-1M}} &
all     & 0.1584 & 0.3449 & 0.1058 & 0.1779 & 0.3770 & 0.1218 & 0.1779 & 0.3770 & 0.1218 \\
\cmidrule(lr){2-11}
& $\mathbb{I}_{\text{f}}$   & \textbf{0.1714} & \textbf{0.3707} & \textbf{0.1151} & \textbf{0.1923} & \textbf{0.4025} & \textbf{0.1331} & \textbf{0.1923} & \textbf{0.4025} & \textbf{0.1331} \\
& $\mathbb{I}_{\text{l}}$ & 0.0846 & 0.1980  & 0.0530 & 0.0958 & 0.2323 & 0.0573 & 0.0958 & 0.2323 & 0.0573 \\
& \textit{Impr.} & \textit{102.6\%} & \textit{87.22\%} & \textit{117.17\%} & \textit{100.73\%} & \textit{73.27\%} & \textit{132.29\%} & \textit{100.73\%} & \textit{73.27\%} & \textit{132.29\%} \\
\cmidrule(lr){2-11}
& $\mathbb{S}_{\text{u}}$    & \textbf{0.1958} & \textbf{0.4145} & \textbf{0.1340} & \textbf{0.2171} & \textbf{0.4501} & \textbf{0.1511} & \textbf{0.2001} & \textbf{0.4222} & \textbf{0.1374} \\
& $\mathbb{S}_{\text{n}}$ & 0.1024 & 0.2405 & 0.0636 & 0.1191 & 0.2674 & 0.0778 & 0.1030 & 0.2363 & 0.0658 \\
& \textit{Impr.} & \textit{91.21\%} & \textit{72.35\%} & \textit{110.69\%} & \textit{82.28\%} & \textit{68.32\%} & \textit{94.22\%} & \textit{94.27\%} & \textit{78.67\%} & \textit{108.81\%} \\
\midrule
\multirow{7}{*}{\textbf{Gowalla}} &
all     & 0.1214 & 0.1950 & 0.0999 & 0.0982 & 0.1639 & 0.0791 & 0.1310 & 0.2090 & 0.1082 \\
\cmidrule(lr){2-11}
& $\mathbb{I}_{\text{f}}$   & \textbf{0.1502} & \textbf{0.241} & \textbf{0.1235} & \textbf{0.1207} & \textbf{0.1998} & \textbf{0.0977} & \textbf{0.1522} & \textbf{0.2406} & \textbf{0.1261} \\
& $\mathbb{I}_{\text{l}}$ & 0.0821 & 0.1376 & 0.0661 & 0.0490 & 0.0856 & 0.0384 & 0.0869 & 0.1436 & 0.0703 \\
& \textit{Impr.} & \textit{82.95\%} & \textit{75.15\%} & \textit{86.84\%} & \textit{146.33\%} & \textit{133.41\%} & \textit{154.43\%} & \textit{75.14\%} & \textit{67.55\%} & \textit{79.37\%} \\
\cmidrule(lr){2-11}
& $\mathbb{S}_{\text{u}}$    & \textbf{0.1466} & \textbf{0.2341} & \textbf{0.1208} & \textbf{0.1215} & \textbf{0.2013} & \textbf{0.0983} & \textbf{0.1568} & \textbf{0.2503} & \textbf{0.1291} \\
& $\mathbb{S}_{\text{n}}$ & 0.1026 & 0.1657 & 0.0842 & 0.0807 & 0.1360 & 0.0647 & 0.1118 & 0.1782 & 0.0925 \\
& \textit{Impr.} & \textit{42.88\%} & \textit{41.28\%} & \textit{43.47\%} & \textit{50.56\%} & \textit{48.01\%} & \textit{51.93\%} & \textit{40.25\%} & \textit{40.46\%} & \textit{39.57\%} \\
\bottomrule
\end{tabular}
\end{table}

In subsection \ref{sec:2.2}, we demonstrate that uniform sequences and frequent items consistently perform better across various datasets. In subsection \ref{sec:2.3}, we further validate this by demonstrating that, regardless of the partitioning thresholds, uniformity and frequency consistently lead to better performance.

\subsection{Symbol Description} \label{sec:2.1}

We distinguish the uniformity and non-uniformity of sequences by adopting the classification method proposed by TiCoSeRec \cite{dang_uniform_2023}, which evaluates and ranks all sequences by calculating the variance of time intervals. Sequences with smaller variances are considered more uniform. Based on this, sequences are divided into two subsets: \(\mathbb{S}_{\text{u}}\) and \(\mathbb{S}_{\text{n}}\). The former includes sequences with consistent time intervals, while the latter contains sequences with significant fluctuations in intervals. Similarly, we rank each item based on the frequency of its occurrence across all user interactions. Define \(\mathbb{I}_{\text{f}}\) as the set of frequently occurring items and \(\mathbb{I}_{\text{l}}\) as the set of less-frequently occurring items.

\subsection{Generality Analysis} \label{sec:2.2}

\subsubsection{Task}
In this experiment, we aim to investigate the comparative recommendation performance on uniform versus non-uniform sequences as well as frequent versus less-frequent items, within the context of different datasets. To achieve balance and fairness, we ensured that subsets \(\mathbb{S}_{\text{u}}\) and \(\mathbb{S}_{\text{n}}\), as well as \(\mathbb{I}_{\text{f}}\) and \(\mathbb{I}_{\text{l}}\), were balanced by equating the interaction numbers as much as possible. Following this division criterion, we assigned "uniformity" and "frequency" labels to each interaction sequence and item, recording the overall evaluation results of the model and the experimental outcomes for data with different labels.

\subsubsection{Experimental Configuration}
TiCoSeRec \cite{dang_uniform_2023} has already demonstrated on several Amazon datasets and Yelp that uniform sequences significantly outperform non-uniform sequences. Here, we extend these findings to both frequent and less-frequent items by testing on two additional datasets,  MovieLens 1M (ML-1M) \cite{harper2015movielens} and Gowalla \cite{cho2011friendship}. The ML-1M dataset, a publicly available movie ratings database, comprises 999,611 ratings from 6,040 users on 3,416 movies, with a sparsity of 95.16\%. The Gowalla dataset, representing check-in data from a location-based social network, contains 6,442,892 check-ins at 1,280,970 unique locations by 107,093 users, with a sparsity of 99.99\%. We utilized three classical sequential recommendation baselines—SASRec \cite{kang_self-attentive_2018}, BERT4Rec \cite{sun_bert4rec_2019}, and LightSANs \cite{fan_lighter_2021} for our analysis. The evaluation metrics include Normalized Discounted Cumulative Gain (NDCG), Hit Rate (HR), and Mean Reciprocal Rank (MRR) at top 20. The evaluation strategy employed is full ranking, which involves evaluating the model on the entire set of items.
\subsubsection{Results Analysis}
Table \ref{table:1} shows the performance of various baselines across two datasets, comparing uniform and non-uniform sequences, as well as frequent and less-frequent items. In the table, "all" represents results tested on the entire dataset, while \(\mathbb{S}_{\text{u}}\) and \(\mathbb{S}_{\text{n}}\), along with \(\mathbb{I}_{\text{f}}\) and \(\mathbb{I}_{\text{l}}\), represent results tested on these specific subsets. The experimental results show that performance on subsets \(\mathbb{S}_{\text{u}}\) and \(\mathbb{I}_{\text{f}}\) is the best, also "all" exceed those on \(\mathbb{S}_{\text{n}}\) and \(\mathbb{I}_{\text{l}}\). For the Gowalla dataset, the Bert4Rec model shows up to a 146.33\% improvement in NDCG@20 when predicting \(\mathbb{I}_{\text{f}}\) instead of \(\mathbb{I}_{\text{l}}\). Similarly, LightSANs improves by up to 94.27\% in NDCG@20 for the ML-1M dataset when transitioning from $\mathbb{S}_{\text{n}}$ to $\mathbb{S}_{\text{u}}$. This phenomenon, where performance on \(\mathbb{I}_{\text{f}}\) substantially exceeds that on \(\mathbb{I}_{\text{l}}\), corroborates the hypothesis that frequent items, benefiting from a larger volume of interaction data, are more predictable. Additionally, models generally exhibit superior performance on \(\mathbb{S}_{\text{u}}\) compared to \(\mathbb{S}_{\text{n}}\), suggesting that models more effectively learn from stable user preferences present in uniform sequences.

\subsection{Invariance Analysis}\label{sec:2.3}
We further explore the impact of different partitioning ratios on model performance using the ML-1M dataset. Specifically, we analyze the effects of varying the ratios for both \(\mathbb{S}_{\text{u}}\) and \(\mathbb{S}_{\text{n}}\) and \(\mathbb{I}_{\text{f}}\) and \(\mathbb{I}_{\text{l}}\) using three classical baseline models.
\begin{figure}
    \centering
    \includegraphics[width=0.95\columnwidth]{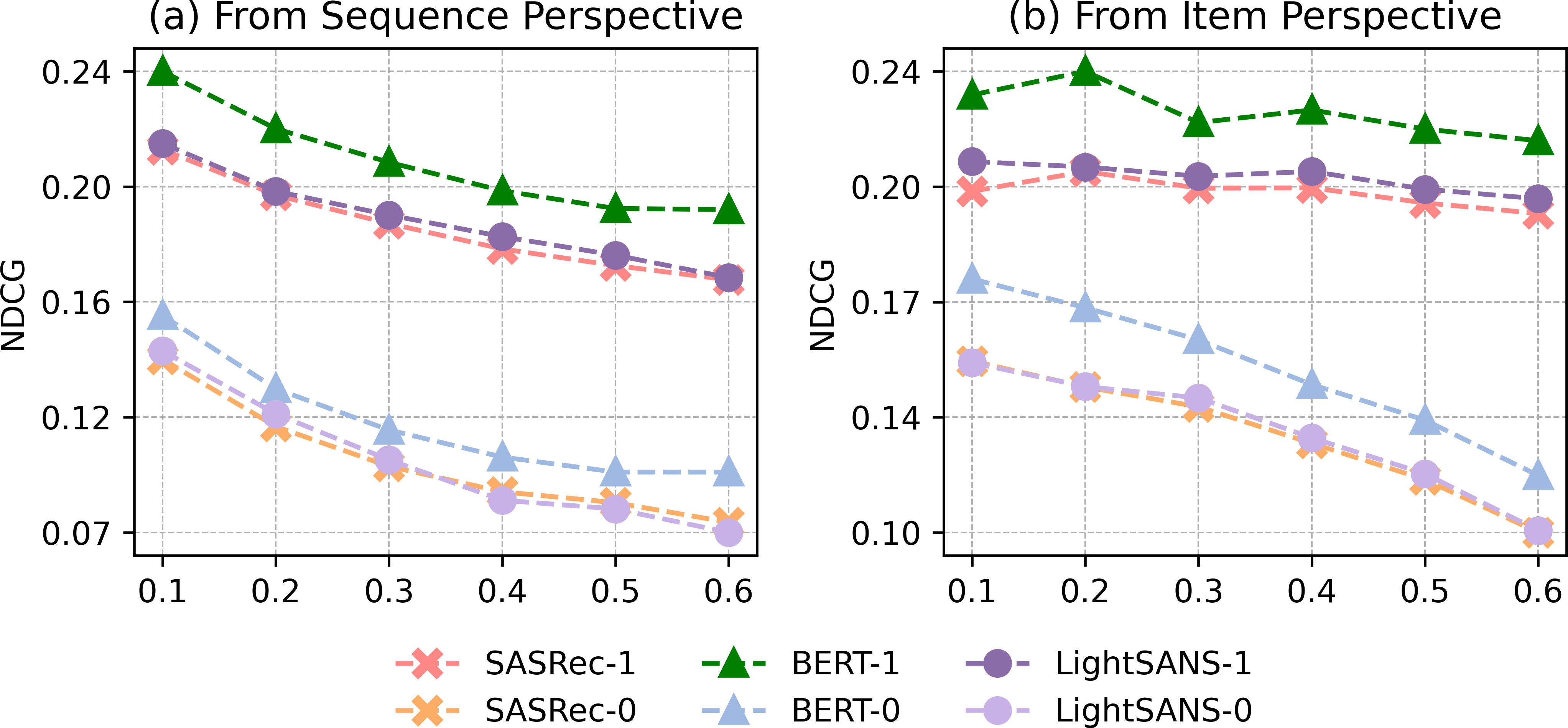}
    \caption{The performance of models under different subset partition ratios, with the X-axis representing the percentage of data classified as uniform and frequent.}
    \label{fig:fig2}
\end{figure}

Figure \ref{fig:fig2}a displays the experimental results on \(\mathbb{S}_{\text{u}}\) and \(\mathbb{S}_{\text{n}}\). In this figure, the "-1" suffix attached to each model indicates the performance on the \(\mathbb{S}_{\text{u}}\), whereas the "-0" suffix indicates the performance on the \(\mathbb{S}_{\text{n}}\). Figure \ref{fig:fig2}b presents the results on \(\mathbb{I}_{\text{f}}\) and \(\mathbb{I}_{\text{l}}\), where "-1" and "-0" similarly denote the performance on \(\mathbb{I}_{\text{f}}\) and \(\mathbb{I}_{\text{l}}\), respectively. The performance trends on MRR@20 and HR@20 are very similar to those observed with NDCG@20. 

The results indicate a noticeable decline in the performance of sequential recommendation models as the partitioning thresholds shift from uniform to non-uniform sequences and from frequent to less-frequent items. This trend highlights the models' sensitivity to the variability in user behavior patterns and item frequencies.

\section{Methodology} \label{sec:3}

\begin{figure}
    \centering
    \includegraphics[width=\columnwidth]{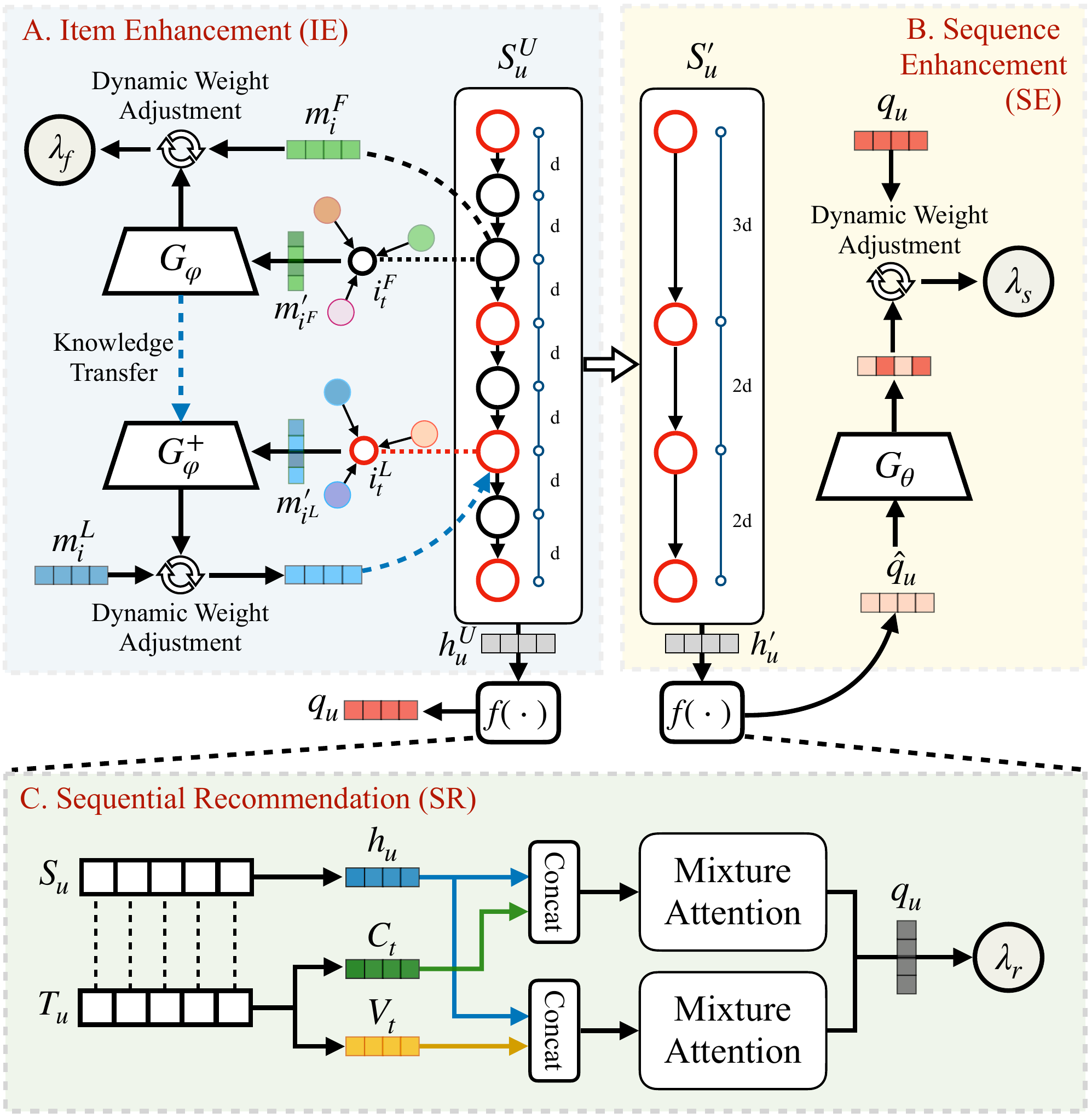}
    \caption{Overview framework of Item Enhancement (A), Sequence Enhancement (B), and Multidimensional Time Modeling in Sequential Recommendation (C), using a uniform sequence as an example.}
    \label{fig:overall}
\end{figure}

This section provides a detailed exposition of UniRec. First, we address the dual enhancement architecture, which comprises the sequences branch (subsection \ref{sec:3.2}) and the items branch (subsection \ref{sec:3.3}). Subsequently, a Multidimensional Time mixture attention module (subsection \ref{sec:3.4}) is designed to accommodate different uniformity sequences. Lastly, subsection \ref{sec:3.5} describes the inference process of the model. Figure \ref{fig:overall} illustrates the overall architecture of the UniRec framework.

\subsection{Problem Formulation}

Let \(\mathcal{U}\) denote the set of all users and \(\mathcal{I}\) represent the set of all items. For each user \(u \in \) \(\mathcal{U}\), we formulate the interactions in chronological order, expressed as \(\mathcal{S}_u^{\text{s-type}} = (i_1^{\text{i-type}}, \dots, i_t^{\text{i-type}}, \dots, i_N^{\text{i-type}})\). Here, \(i_t^{\text{i-type}} \in \mathcal{I}\) specifies the item with which the user interacted at timestamp \(t\). The term "s-type" distinguishes a sequence as uniform or non-uniform, denoted as \(\mathcal{S}_u^{\text{U}}\) and \(\mathcal{S}_u^{\text{N}}\); "i-type" identifies an item as frequent or less-frequent as \(i_t^{\text{F}}\) and \(i_t^{\text{L}}\), respectively. \(N\) signifies the sequence length, which is fixed. For sequences shorter than \(N\), we employ the padding operation to fill the missing parts and for those longer than $N$ we truncate the excess part. Define \( M_I \in  \mathbb{R}^{\mathcal{I} \times d}\)  as a learnable matrix of all items' embedding, $d$ is a positive integer denoting the latent dimension. By performing a lookup table operation on \( M_I \), we can retrieve every single item embedding \( m_i \in  \mathbb{R}^{d}\), to form the user embedding \( h_u =[m_1, \ldots, m_t, \ldots, m_N] \in \mathbb{R}^{N \times d} \).

\subsection{Sequence Enhancement} \label{sec:3.2}

Sequences with smaller variances are considered
more uniform and sequences are divided into two subsets: \(\mathbb{S}_u\) and \(\mathbb{S}_n\). Each sequence is classified based on a predefined time variance threshold into either \(\mathcal{S}_u^{\text{U}}\) or \(\mathcal{S}_u^{\text{N}}\), where  \(\mathcal{S}_u^{\text{U}} \in \mathbb{S}_u\) and \(\mathcal{S}_u^{\text{N}} \in \mathbb{S}_n\). Similarly, item is categorized based on their frequency of occurrence in interactions into \(i_t^{\text{F}}\) or \(i_t^{\text{L}}\), where \(i_t^{\text{F}} \in \mathbb{I}_{\text{f}}\) and \(i_t^{\text{L}} \in \mathbb{I}_{\text{l}}\). For each uniform sequence \(\mathcal{S}_u^{\text{U}}\), we generate a corresponding non-uniform sub-sequence \(\mathcal{S}'_u\) to emulate the irregular patterns observed in real-world datasets, thereby enhancing the capability to model complex user behaviors. The generation process retains all items from \(\mathbb{I}_{\text{l}}\) within \(\mathcal{S}_u^{\text{U}}\), and if the count of  \( i_t^{\text{L}} \in \mathcal{S}_u^{\text{U}} \) is fewer than \(M\), additional \(i_t^{\text{F}}\) are randomly sampled from \(\mathcal{S}_u^{\text{U}}\), where \(M\) is the hyper-parameter of the minimum length of  \(\mathcal{S}'_u\):
\begin{equation}
  \mathcal{S}'_u = \begin{cases} 
     \text{if } \text{count}(\mathcal{S}_u^{\text{U}}, \mathbb{I}_{\text{l}}) <  M:\\ 
\hspace{0.5cm} \{i_t^{\text{L}}: i_t^{\text{L}} \in \mathcal{S}_u^{\text{U}}\} \cup \{\text{Sampled } i_t^{\text{F}} \in \mathcal{S}_u^{\text{U}}\}
     \\
     \text{otherwise}: 
    \\ \hspace{0.5cm} \{i_t^{\text{L}}: i_t^{\text{L}} \in \mathcal{S}_u^{\text{U}}\}
  \end{cases}
\end{equation}
the variance of time intervals increases from the sequence \(\mathcal{S}_u^{\text{U}}\) to \(\mathcal{S}'_u\), and there is a substantial rise in the relative composition of \(i_t^{\text{L}}\) within \(\mathcal{S}'_u\).

We utilize \(\mathcal{S}_u^{\text{U}}\) to enhance the model's learning capability with respect to \(\mathcal{S}'_u\). First, we generate the initial embeddings for \(\mathcal{S}_u^{\text{U}}\) and \(\mathcal{S}'_u\), denoted as \(h_u^U \in \mathbb{R}^{N \times d}\) and \(h'_u \in \mathbb{R}^{N \times d}\) respectively. For each sequence, we employ a sequence encoder \( f(\cdot) \), which is the sequential recommendation modeling process: 
\begin{equation}
    q_u = f(h_u^U),\
    \hat{q_u} = f(h'_u)
\end{equation}
where \( q_u \in \mathbb{R}^{N \times 2d} \) and \( \hat{q_u} \in \mathbb{R}^{N \times 2d} \) are the representations for \(\mathcal{S}_u^{\text{U}}\) and \(\mathcal{S}'_u\). The specifics of \( f(\cdot) \) will be detailed in subsection \ref{sec:3.4}. Next, the objective is to bring \(q_u\) and \(\hat{q_u}\) as close as possible in the feature space to enhance the model’s ability to handle the temporal dynamics of non-uniform sequences, thereby minimizing \(\tilde{x}\) through a generative model $\boldsymbol{G}_\theta$, which consists of a feed-forward layer: 
\begin{equation}
    \tilde{x} = q_u-\boldsymbol{G}_\theta (\hat{q_u})
\end{equation}
Meanwhile, a curriculum learning strategy is adopted, which mimics the human learning process: from simple to complex. This strategy gradually increases the training samples' complexity. Specifically, the model initially learns predominantly from more uniform sequences, while sequences with more complex user interest drifts are introduced later in the training. This process is managed with a dynamically weighted loss function \(\lambda_{s}\) guiding the progression:
\begin{equation}
    \lambda _{s} =w_s || \tilde{x} ||^2 
\end{equation}
\begin{equation}
    w_s=  \sin\left(\frac{\pi}{2} \cdot \frac{e - e_b}{e_{all}} + \frac{\pi}{2} \cdot \frac{V_{max} - V_u}{V_{max}-V_{min}}\right) 
\end{equation}
where \( w_s \) represents a dynamic weight coefficient, \( e \) denotes the current epoch number, \( e_b \) denotes the epoch at which this loss function starts to contribute to the training process, and \( e_{all} \) denotes the total number of training epochs. For each \(\mathcal{S}_u^{\text{U}}\in \mathbb{S}_u\), the variance of the time intervals is defined as \(V_u\). \(V_{max}\) is the maximum time interval variance among all sequences, while \(V_{min}\) is the minimum. This design allows \( w_s \) to dynamically change its value during the training process based on the uniformity of sequences and phases of training progress. This task serving as an auxiliary task, parallel to the main task of sequential recommendation, specifically enhances the model's performance on  \(\mathcal{S}'_u\), thereby implicitly improving the model's adaptability and prediction accuracy on \(\mathbb{S}_n\).

\subsection{Item Enhancement} \label{sec:3.3}
Given that the generated \(\mathcal{S}'_u\) are predominantly composed of \(i_t^{\text{L}}\), together with a general prevalence of \(i_t^{\text{L}}\) in \(\mathcal{S}_u^{\text{N}}\), enhancing model performance on \(i_t^{\text{L}}\) has become critical. The proposed item enhancement approach operates from two aspects: utilizing the information from neighboring items and leveraging the knowledge transferred from \(i_t^{\text{F}}\) to \(i_t^{\text{L}}\). Leveraging neighbors for enhancement involves two steps: candidate neighbor generation and representation aggregation.

Initially, the candidate neighbor generation process is conducted for each item. For each center item \(i_c \in \mathcal{I}\), a potential candidate neighbor set \(\mathbb{N}_{i_c}\) is identified. A bunch of score \(s(i_c, j)\) is calculated for \(i_c\) against every other item \(j\) (where \(j \in \mathcal{I} \setminus \{i_c\}\)). These scores are then ranked, and the items with higher scores are chosen to constitute the neighbor set \(\mathbb{N}_{i_c}\). \(s(i_c, j)\) integrated three factors: the temporal interval \(T\) between \(i_c\) and \(j\), the popularity \(H\) of item \(j\), and the similarity \(S\) between \(i_c\) and \(j\). Both \(H\) and \(S\) are normalized to ensure consistency in the scoring mechanism. \( s(i_c, j) \) is defined as:
\begin{equation}
    s(i_c, j) = g(T) + \phi(T, H) + \phi(T, S) 
\end{equation}
\begin{equation}
    g(T) = \frac{1}{1 + \log(1 + T)} 
\end{equation}
\begin{equation}
    \phi(T, x) = \frac{T + \Theta}{e^{(T + \Theta)/\Gamma x}}
\end{equation}    
where \(\Theta\) and \(\Gamma\) are constants, determined based on dataset specifics. As \( T \) increases, \( g(T) \) gradually decreases. Similarly, an increase in \(T\) or a decrease in \(x\) results in a lower value of \(\phi(T, x)\). This scoring framework adeptly manages the temporal dynamics among items, accounting for factors such as the popularity and similarity of potential neighboring items. In each training batch, \( K \) neighbors are randomly sampled from \(\mathbb{N}_{i_c}\), where \(K\) is a hyper-parameter. 

Then we aggregate these \( K \) candidate neighbors to enhance \( i_c \) with a simple attention mechanism. We generate the initial embedding for \( i_c \), denoting as \( m_c \in  \mathbb{R}^{d}\ \),  as well as the embedding \( m_o \in  \mathbb{R}^{d}\) for these \( K \) neighbors, \(o \in \{1, 2, \ldots, K\} \). The aggregation process is as follows: 
\begin{equation}
       m_n = \sum_{k=1}^K \frac{\exp(m_c^T m_k)}{\sum_{j=1}^K \exp(m_c^T m_j)}
\end{equation}
\( m_n \) represents the aggregated embedding from the neighbors. We then concatenate \( m_n \) and \( m_c \) to form the updated representation \( m_c' = [m_c \parallel m_n]\in\mathbb{R}^{2d}\), where \( || \) denotes the concatenation operation. As a result, \( m_c' \) contains more information related to \( i_c \) than \( m_c \).

Meanwhile, to enable \( i_t^{\text{L}} \in \mathbb{I}_{\text{l}} \) to better utilize the related information from \(\mathbb{N}_{i_c}\), we transfer the knowledge learned from \( \mathbb{I}_{\text{f}} \) on neighbor aggregation representation to \( \mathbb{I}_{\text{l}} \). Define the embedding of \( i_t^{\text{F}} \) obtained from \( M_I \) as \( m_i^F \). Define the updated embedding \( m_c'\) of \( i_t^{\text{F}} \) as \( m'_{i^F} \). We train the aggregation mechanism on \( i_t^{\text{F}} \) by minimizing the following loss function:

\begin{equation}
    \lambda _{f} =w_i ||m_i^F - \boldsymbol{G}_\varphi ( m'_{i^F} )||^2 
\end{equation}
\begin{equation}
    w_i=  \sin\left(\frac{\pi}{2} \cdot \frac{e - e_b}{e_{all}} + \frac{\pi}{2} \cdot \frac{F - F_{min}}{F_{max}-F_{min}}\right) 
\end{equation}
where \(\boldsymbol{G}_\varphi\) is a fully connected layer that aligns the dimensions of \(m'_{i^F}\) and \(m_i^F\) to be consistent. \(w_i\) is a dynamic parameter used to adjust the magnitude of the loss function across different items. \(F\) represents the frequency score of the current item across all interactions. \(F_{\text{min}}\) is the minimum \(F\) of \(i_t^{\text{F}} \in \mathbb{I}_{\text{f}}\), while \(F_{\text{max}}\) is the maximum. A curriculum learning strategy, analogous to the sequence branch, is also employed. In the initial training phase, high-frequency items are prioritized, with a gradual shift towards less-frequent items in the later stages.

Finally, update the embeddings of all \( i_t^{\text{L}} \) after a certain epoch $e_t$ of training by minimizing the following loss:
\begin{equation}
   \lambda _{l} = \eta|| m_i^L -  \boldsymbol{G}_\varphi^+(m'_{i^L})||^2
\end{equation}
\begin{equation}
    \eta = \sin (\frac{\pi }{2} \cdot \frac{e-e_t}{e_{all}}  )
\end{equation}
where \( m_i^L \) is the representation of \( i_t^{\text{L}} \) obtained from \( M_I \), \( m'_{i^L} \) is the updated representation \( m_c'\) of \( i_t^{\text{L}} \), and \(\eta\) is a parameter that dynamically increases with the increase of the training epoch. \(\boldsymbol{G}_\varphi^+\) represents the \(\boldsymbol{G}_\varphi\) trained after $(e-e_b)$ epochs and is static. By refining \(i_t^{\text{L}}\) representation through the auxiliary task before the main task training, the accuracy and performance of the model concerning \(i_t^{\text{L}}\) are improved.

\subsection{Multidimensional Time Modeling} \label{sec:3.4}
Given the varying dependencies on temporal information, where \(\mathcal{S}_u^{\text{U}}\) has a lower reliance on time and \(\mathcal{S}_u^{\text{N}}\) requires richer temporal details, we propose a multidimensional time modeling module to accommodate these differing needs. As demonstrated in subsection \ref{sec:4.6}, utilizing time interval information is more effective for \(\mathcal{S}_u^{\text{U}}\), while employing comprehensive temporal context proves more effective for \(\mathcal{S}_u^{\text{N}}\). Therefore, we design this module to better leverage the appropriate temporal information. 

For each \(\mathcal{S}_u\) we define its corresponding timestamp sequence as \(\mathcal{T}_u = (t_1, t_2, \dots, t_N)\). The corresponding time interval sequence is defined as \(\mathcal{T}_{\text{intv}} = (\tau_1, \tau_2, \dots, \tau_{N-1})\), where each \(\tau_k = t_{k+1} - t_k\) denotes the interval between the \( k^{\text{th}} \) and \((k+1)^{th}\) interactions. Each \(\tau_k\) is encoded by an embedding matrix, resulting in a time interval embedding \(v_k \in \mathbb{R}^{d} \). For temporal context modeling, we adopted the approach proposed by Xu et al. \cite{xu2019self}, which specifically uses a self-attention mechanism based on time representation learning, and models temporal information such as year, month, and day separately. Subsequently, this information is aggregated through a linear layer to form the final temporal context embedding $c_i \in \mathbb{R}^{d}$ for each interaction $i$. In a word, for each $S_u$, we obtain its item sequence embedding \(h_u \in \mathbb{R}^{N \times d}\), along with the temporal context representation \(C_t = [c_1, c_2, \ldots, c_N] \in \mathbb{R}^{N \times d}\), and the time interval embeddings \(V_t = [0, v_1, v_2, \ldots, v_{n-1}] \in \mathbb{R}^{N \times d}\), 0 represents a \(1 \times d\) zero vector. 

Next, recognizing that sequences with different uniformity require varying levels of temporal information, we integrate \(h_u\) with \(C_t\) and \(V_t\) respectively using a mixture attention mechanism. This serves as the sequence encoder \( f(\cdot) \), generating \(q_u\), the embedding of the user \(u\)'s interaction sequence, tailored to the specific needs of each sequence. Integrate \( h_u \) with \( C_t \) and \( V_t \) in the same way, taking the application of mixture attention on \( h_u \) and \( C_t \) as an example. First, concatenate \( h_u \) and \( C_t \) to obtain the initial embedding of a sequence as \( e_u = h_u \,||\, C_t \). Next, we preprocess the input \( X \) for mixture attention, which is defined as \( X = e_u + P \), where \( P \in \mathbb{R}^{N \times 2d}\) is the position encoding matrix. The mixture attention mechanism can be mathematically described as: 
\begin{equation}
    \text{MixATT}(X) = \text{FFL}(\text{SAL}(X)) 
\end{equation}
\begin{equation}
    \text{FFL}(X) = \text{ReLU}(XW_F + b_F)W_{F'} + b_{F'} 
\end{equation}
\begin{equation}
    \text{SAL}(X) = \text{Concat}(H_1, \ldots, H_H) 
\end{equation}
where \(\text{MixATT}(X)\) represents a composite model that integrates a self-attention mechanism \(\text{SAL}(X)\) and a feed-forward layer \(\text{FFL}(X)\). \(\text{FFL}\) involves two linear transformations with weight matrices \(W_F\) and \(W_{F'}\), and bias terms \(b_F\) and \(b_{F'}\). \(\text{SAL}\) combines the outputs \(H_j\) from each attention head \(j \in \{1, \ldots, H\}\). Each \(H_j\) is given by \(\text{softmax}(A_j / \sqrt{d_V}) W_j^O\), where \(\sqrt{d_V}\) is a scaling factor to stabilize learning, and \(W_j^O\) is the output projection matrix for the \(j^{th}\) head. \( A_j \) is the attention score matrix proposed by Viet-Anh Tran et al. \cite{tran_attention_2023}, combining Gaussian distribution to mix two types of input data. \( A_j = \sum_{k \in \{m, c\}} p_{kj} \mathcal{N}(A; Q_k^T, \sigma^2 I) \) is approximated by a mixture model. The non-negative mixture weights \( p_{kj} \) sum to one, indicating the contribution of each context type. \( Q_k \) is obtained by projecting the input context \( X_k \) using matrix \( W_k \). The Gaussian distribution's variance parameter is \(\sigma^2\), and \( I \) is the identity matrix.

The loss function for the recommendation task can be defined as follows:
\begin{equation}
      \lambda _{r}=q_u n_i^{\mathsf{T}}
\end{equation}
where $q_u$ is the output of the FFL and $n_i=[m_i||c_i]$ is the embedding of the next item to be predicted. Similarly, the mixture attention mechanism is also applied to \( h_u \) and \( V_t \). The outputs processed through the mixture attention mechanism, are mutually supervised within a multi-task learning framework.

\subsection{Inference Process} \label{sec:3.5}

\begin{figure}
    \centering
    \includegraphics[width=\columnwidth]{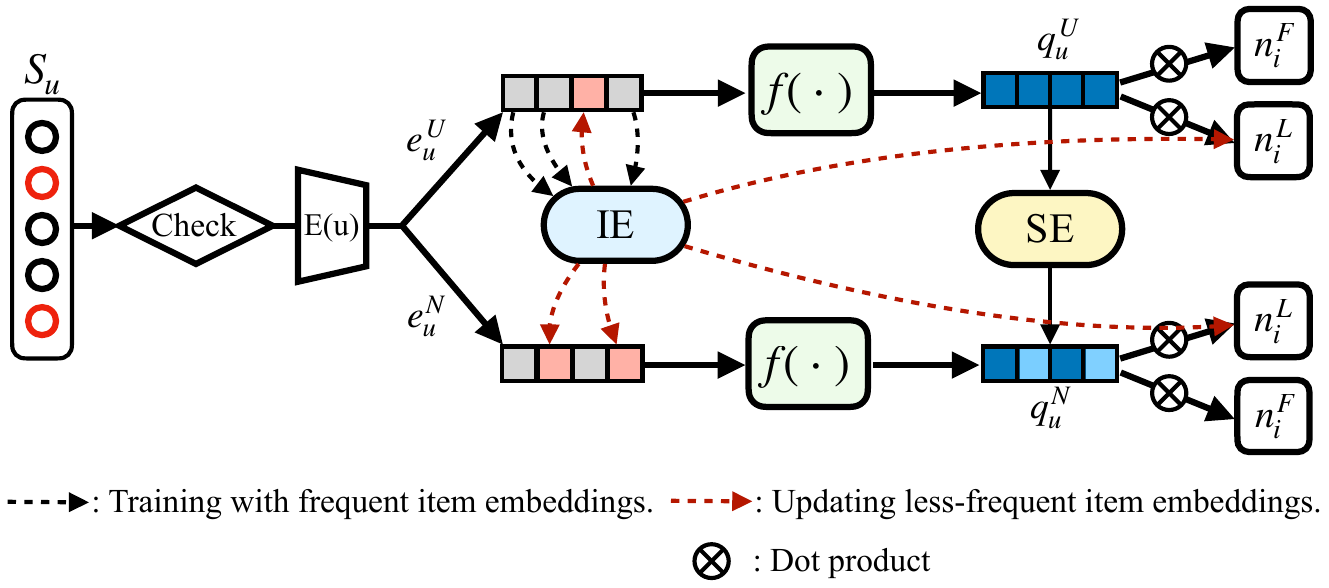}
    \caption{Overview of inference phase.}
    \label{fig:inference}
\end{figure}

\begin{table*}[htbp]
\setlength{\tabcolsep}{3.1 pt} 
\setlength{\extrarowheight}{2pt} 
\footnotesize 
\belowrulesep=1.5pt 
\aboverulesep=1.5pt 
\caption{Performance comparison over four datasets. Numbers in bold indicate the best performance, those underlined denote the second best, and numbers marked with an asterisk represent the third best. Models marked with $^{\dagger}$ are data-augmented methods based on SASRec.}
\label{tab:2}
\begin{tabular}{c|c|ccccccc|ccccc|c}
\toprule
\multirow{2}{*}{\textbf{  Dataset  }} & \multirow{2}{*}{\textbf{Metric}} & \multicolumn{7}{c|}{\textbf{Non-Time-Aware}} & \multicolumn{6}{c}{\textbf{Time-Aware}} \\
\cmidrule(lr){3-9} \cmidrule(lr){10-15}
& & GRU4Rec & Caser & STAMP & SASRec & BERT4Rec & LightSANs & DuoRec$^{\dagger}$ & TiSASRec & Meantime & TiCoSeRec$^{\dagger}$ & FEARrec & MOJITO & \textbf{ UniRec } \\
\midrule
\multirow{3}{*}{\textbf{ML-1M}}
& \textbf{NDCG@10} & 0.5758 & 0.5447 & 0.5302 & 0.5801  & 0.5658 & 0.5671 & 0.5816 & 0.5849*  & 0.5804  & 0.5732 & 0.5515 & \underline{0.5929} & \textbf{0.6261} \\
& \textbf{HR@10}   & 0.7856 & 0.7692 & 0.7444 & 0.8098* & 0.7780 & 0.7719 & 0.7971   & 0.7893  & 0.8098* & 0.7904 & 0.7594 & \underline{0.8197} & \textbf{0.8347} \\
& \textbf{MRR@10}  & 0.5093 & 0.4751 & 0.4629 & 0.5149  & 0.4987 & 0.5021 & 0.5044 & \underline{0.5205} & 0.5079  & 0.4967 & 0.4856 & 0.5157*      & \textbf{0.5613} \\ \midrule
\multirow{3}{*}{\textbf{Beauty}}
& \textbf{NDCG@10} & 0.3014 & 0.2805 & 0.2809 & 0.2944 & 0.3140 & 0.3349*   & 0.3123   & 0.2908 & 0.3201 & 0.3188 & 0.3382  & \underline{0.3392} & \textbf{0.3693} \\
& \textbf{HR@10}   & 0.4593 & 0.4394 & 0.4225 & 0.4418 & 0.4629 & 
 0.5042*  &   0.4828   & 0.4323 & 0.4655 & 0.4862 & 0.4863  & \underline{0.5087} & \textbf{0.5313} \\
& \textbf{MRR@10}  & 0.2525 & 0.2313 & 0.2372 & 0.2490 & 0.2682 & \underline{0.2986} & 0.2733 & 0.2471 & 0.2753 & 0.2841 & 0.2922* & 0.2865       & \textbf{0.3201}\\ \midrule
\multirow{3}{*}{\textbf{Books}}
& \textbf{NDCG@10} & 0.5811 & 0.5356 & 0.4816 & 0.6072  & 0.5616 & 0.6049 & 0.5942 & 0.6045       & 0.6073* & 0.5875 & 0.5745 & \underline{0.6171} & \textbf{0.6309} \\
& \textbf{HR@10}   & 0.7951 & 0.7674 & 0.7078 & 0.8216  & 0.7868 & 0.8176 & 0.8130 & 0.8218*      & 0.8169  & 0.8023 & 0.7994 & \underline{0.8513} & \textbf{0.8617} \\
& \textbf{MRR@10}  & 0.5134 & 0.4626 & 0.4107 & 0.5394* & 0.4905 & 0.5378 & 0.5325 & 0.5357 & 0.5135  & 0.5278 & 0.5036 & \underline{0.5428} & \textbf{0.5752}\\ \midrule
\multirow{3}{*}{\textbf{Toys}}
& \textbf{NDCG@10} & 0.2779 & 0.2173 & 0.2446 & 0.3118 & 0.2327 & \underline{0.3364} & 0.2794 & 0.3224       & 0.3187 & 0.2754 & 0.3121 & 0.3323* & \textbf{0.3609} \\
& \textbf{HR@10}   & 0.4432 & 0.3752 & 0.3886 & 0.4626 & 0.3887 & 0.4934*     & 0.4643  & 0.4766       & 0.4762 & 0.4431 & 0.4655 & \underline{0.5087}  & \textbf{0.5260} \\
& \textbf{MRR@10}  & 0.2270 & 0.1688 & 0.2004 & 0.2652 & 0.1848 & \underline{0.2877} & 0.2687 & 0.2747 & 0.2711 & 0.2421 & 0.2647 & 0.2775* & \textbf{0.3103}\\
\bottomrule
\end{tabular}
\end{table*}

Figure \ref{fig:inference} shows how the integrated components—IE (Item Enhancement), SE (Sequence Enhancement), and \( f(u) \) (Sequential Recommendation)—work together to provide robust and contextually rich recommendations. For a given input sequence \(\mathcal{S}_u\), we first determine whether it is \(\mathcal{S}_u^{\text{U}}\) or \(\mathcal{S}_u^{\text{N}}\). \(\mathcal{S}_u^{\text{U}}\) is initialized with embedding \( e_u^U \), while \(\mathcal{S}_u^{\text{N}}\) is initialized with \( e_u^N \). Within each \(\mathcal{S}_u^{\text{U}}\), \( i_t^{\text{F}} \) are utilized to train \(\boldsymbol{G}_\varphi\) through the loss function \(\lambda_f\) in the IE module. Conversely, for both \(\mathcal{S}_u^{\text{U}}\) and \(\mathcal{S}_u^{\text{N}}\), \( i_t^{\text{L}} \) are updated based on the output from \(\boldsymbol{G}_\varphi^+\) using the loss \(\lambda_l\). After processing the sequence through \( f(u) \), we train its embedding via the primary task loss \(\lambda_r\). The sequence embedding is then refined by the SE module to further enhance the sequence representation using the loss \(\lambda_s\). Finally, the sequence embedding and the embedding of the item to be predicted are scored by calculating their dot product.

\section{Experiment}

\subsection{Experimental Settings}
\subsubsection{Datasets}
In addition to the ML-1M \cite{harper2015movielens} dataset used in section \ref{sec:2}, we also use datasets from e-commerce platforms, including those for books, beauty products, and toys, as detailed below:
\begin{enumerate}
    \item The Amazon Book \cite{he2016ups} dataset consists of 6,275,735 interactions of users rating a book. This dataset includes 79,713 users and 91,465 books, with a density of 0.00086, indicating the sparsity of user-item interactions.
    \item The Amazon Beauty \cite{mcauley2015image} dataset comprises 198,502 interactions involving 22,363 users and 12,101 beauty products, with a density of 0.00073.
    \item The Amazon Toys \cite{mcauley2015image} dataset includes 167,597 interactions from 19,412 users and 11,924 toys, with a sparse density of 0.00072.
\end{enumerate}

For each dataset, we adopt the k-core filtering \cite{sarwar2001item} as a pre-processing step, which iteratively removes users and items whose interactions are fewer than \( k \), until each user and item in the dataset has at least \( k \) interactions. Specifically, for the ML-1M, we set \( k_{\text{item}} = 5 \) and \( k_{\text{user}} = 10 \); for the Beauty and Toy, we set \( k_{\text{item}} = 5 \) and \( k_{\text{user}} = 5 \); and for the Books, the settings are \( k_{\text{user}} = 30 \) and \( k_{\text{item}} = 20 \).

\subsection{Evaluation Settings}
We arrange the dataset in chronological order and allocate the last item as the validation set and the penultimate item as the test set, using the remaining data to construct the training set. To ensure fair evaluation, for each positive item in the test set, we pair it with 100 negative items sampled uniformly, and the model's performance is assessed based on these pairs. We primarily utilize three metrics for performance evaluation based on top-10 recommendation results: NDCG, HR, and MRR. Specifically, NDCG assesses the ranking quality of recommended items, HR measures the presence of at least one relevant item, and MRR evaluates the rank of the top relevant item.

\subsubsection{Comparison Methods}
We conduct a comprehensive comparison of UniRec with 11 baseline models. These include six classic sequential recommendation models: GRU4Rec \cite{jannach_when_2017}, Caser \cite{tang_personalized_2018}, STAMP \cite{liu_stamp_2018}, SASRec \cite{kang_self-attentive_2018}, BERT4Rec \cite{sun_bert4rec_2019}, and LightSANs \cite{fan_lighter_2021}. Additionally, we evaluate five time-aware models: TiSASRec \cite{li_time_2020}, Meantime \cite{cho_meantime_2020}, TiCoSeRec\cite{dang_uniform_2023}, FEARec \cite{du_frequency_2023}, and MOJITO \cite{tran_attention_2023}, all of which leverage temporal information to improve performance.

\subsubsection{Implementation Details}
All models are trained for up to 200 epochs utilizing the Adam optimizer \cite{kingma2014adam}. Early stopping is implemented with a patience threshold of 20 epochs. We assign a value of 64 to the parameter \(d\), utilize a batch size of 512, and set the learning rate to 0.01. The length of the sequence is fixed at 50. Both hyper-parameters \(M\) and \(K\) are set to 3. The mixture attention mechanism is configured with 2 heads. We test the partitioning ratios for uniform and non-uniform users within the range of \{0.3, 0.4, 0.5, 0.6, 0.7, 0.8\}, and for frequent and less-frequent items within the range of \{0.4, 0.5, 0.6, 0.7, 0.8, 0.9\}, across each dataset.

\subsection{Overall Performance} \label{sec:4.2}

\begin{figure*}[ht]
    \centering
    \includegraphics[width=\textwidth]{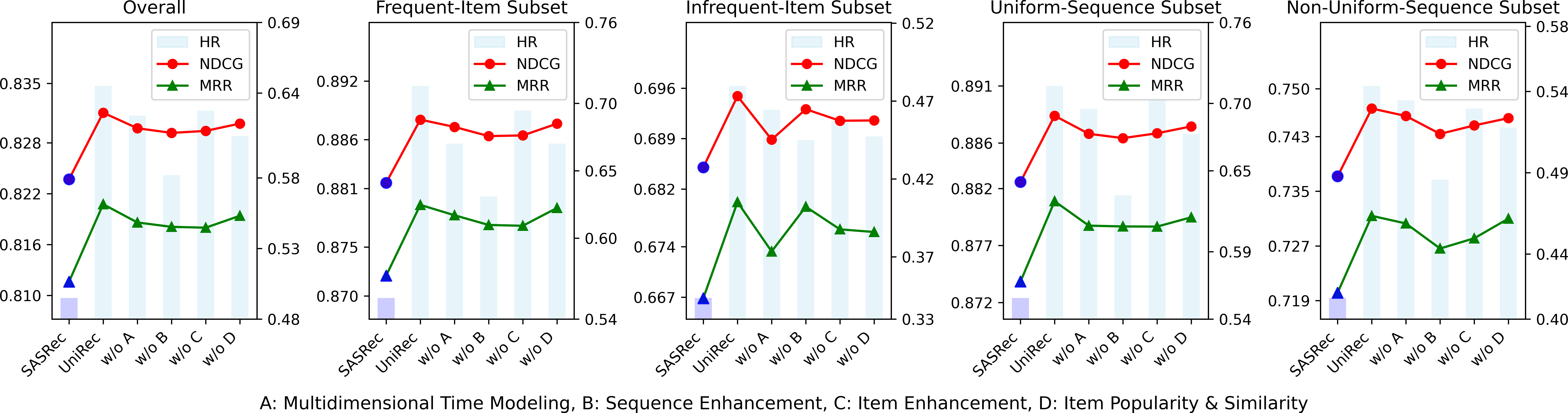}
    \caption{Ablation performance with various enhancements across different subsets from ML-1M.}
    \label{fig:abla}
\end{figure*}

Table \ref{tab:2} presents the experimental results of UniRec and 11 baselines across four datasets, several conclusions can be drawn. First, time-aware models generally outperform non-time-aware sequential recommendation models across various datasets. This highlights the critical importance of incorporating temporal dynamics into the recommendation process, as it substantially enhances the relevance and accuracy of the recommendations. Second, UniRec significantly outperforms other comparative models across all datasets and evaluation metrics, confirming its effectiveness. The bidirectional enhancement strategy for sequences and items adopted by UniRec, along with the multidimensional time modeling, greatly enhances the precision in modeling user interests and item characteristics. For instance, on the ML-1M dataset, UniRec achieves improvements of 3.32\% in NDCG@10 and 4.08\% in MRR@10 compared to the existing SOTA techniques. Third, UniRec demonstrates exceptional performance across datasets with varying sparsity and scale, whether in the lower-sparsity, smaller-scale ML-1M dataset or in the larger, more sparse Amazon datasets. This proves its adaptability and robustness to different levels of sparsity and data sizes. For example, on the Books dataset, UniRec increases MRR@10 by 3.24\%, and on the Beauty dataset, it raises NDCG@10 by 3.01\%. Lastly, compared to TiCoSeRec, which enhances data by improving sequence uniformity, UniRec enhances the utilization of sequence uniformity by incorporating item frequency more effectively. This demonstrates the potential of enhancing sequential recommendations from both perspectives of item frequency and sequence uniformity.

\subsection{Ablation Experiment} \label{sec:4.3}
To understand the impact of various components in our model, we conduct an ablation study. We divide the model into the following parts for evaluation: Multidimensional Time Modeling (A), Sequence Enhancement (B), Item Enhancement (C), and Item Popularity \& Similarity (D). Specifically, w/o A refers to the replacement of multidimensional time modeling with a single-dimensional time modeling structure, utilizing only time interval modeling and disregarding contextual time information. w/o B refers to removing the sequence enhancement task, while w/o C refers to removing the item enhancement task. w/o D refers to excluding the consideration of item popularity and similarity in the item enhancement component, instead selecting candidate neighbors based solely on the time interval of the project. In addition to the overall dataset results, we evaluate performance on several subsets: frequent-item, less-frequent-item, uniform-sequence, and non-uniform-sequence. Using the ML-1M dataset as an example, Figure \ref{fig:abla} shows the evaluation results of SASRec, UniRec, and UniRec without several components across various subsets.

First, UniRec demonstrates significant performance improvements over SASRec across all strategies, particularly in the less-frequent-item and non-uniform-sequence subsets. According to the experimental data, UniRec shows a 9.2\% improvement in MRR@10 over SASRec in the frequent-item subset and an 18.0\% improvement in the less-frequent-item subset. Additionally, in uniform and non-uniform subsets, UniRec achieves a 2.1\% and 4.3\% improvement in HR@10 over SASRec, respectively. These findings indicate that UniRec excels in enhancing performance for less-frequent items and non-uniform sequences.

Secondly, removing each component of the model results in varying degrees of performance degradation, indicating the importance of each component to the overall model performance. Particularly, w/o B leads to the most significant performance drop, particularly reflected in the HR metric, highlighting the effectiveness of the sequence enhancement module. This module not only improves the uniformity of non-uniform sequences but also increases the frequency of less-frequent items, significantly contributing to the accuracy of user interest modeling.

Furthermore, the performance on the frequent-item subset and uniform-sequence subset is consistent with the overall data. However, there are some differences between the less-frequent-item subset and the non-uniform-sequence subset. In the less-frequent-item subset, w/o A shows a significant drop in NDCG@10 and MRR@10, indicating that temporal information has a substantial impact on less-frequent items, as certain less-frequent items are more likely to be interacted with during specific periods. The declines in NDCG@10 and MRR@10 for w/o C and w/o D also demonstrate the effectiveness of these components in modeling less-frequent items. In particular, w/o D underscores the importance of considering item popularity, similarity, and relevance in selecting candidate neighbors to enhance less-frequent items' representations. In the non-uniform-sequence subset, the significant performance drop in w/o B indicates that sequence enhancement indeed improves the model's capability to handle sequences with rich interest drifts.

In summary, Figure \ref{fig:abla} clearly illustrates the contributions of each component to the performance of UniRec, validating the necessity and effectiveness of multidimensional time modeling, sequence enhancement, item enhancement, and item popularity \& similarity in improving the model's recommendation performance.

\begin{figure}[ht]
    \centering
    \includegraphics[width=0.95\columnwidth]{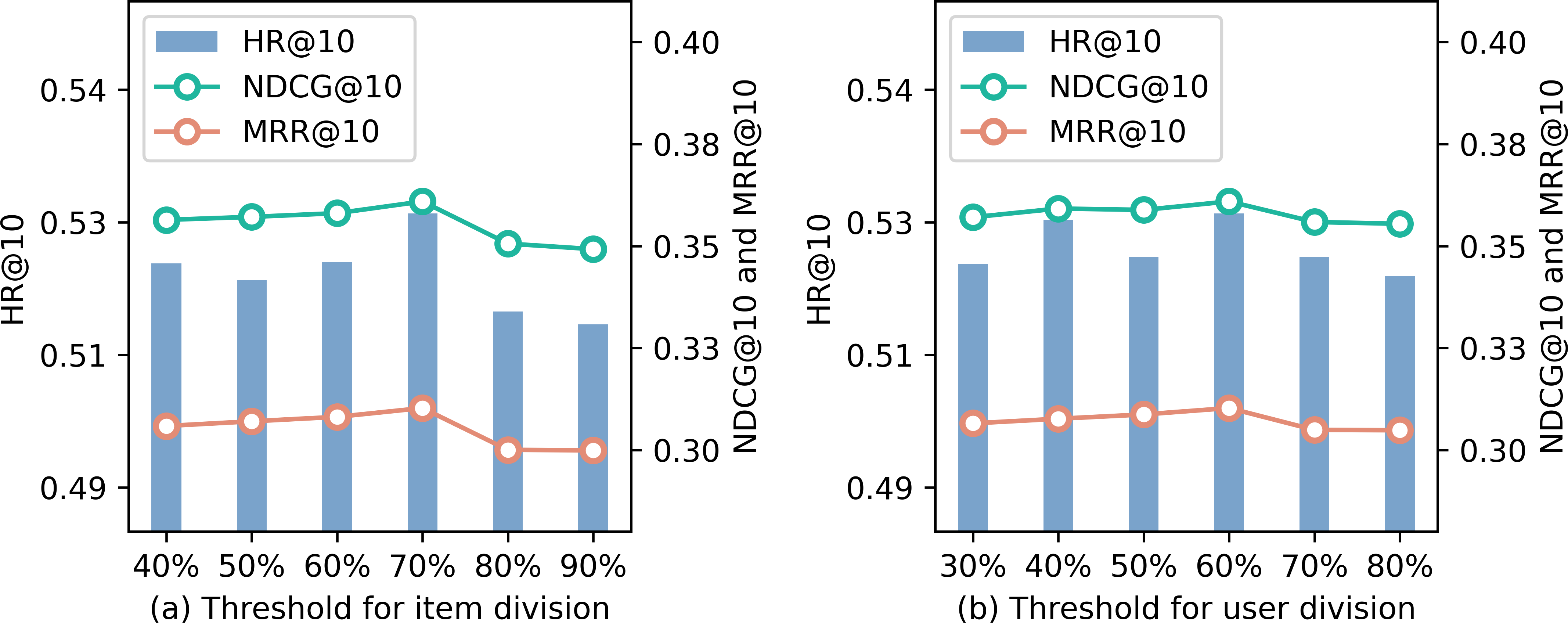}
    \caption{Performance comparison using different partition thresholds for item frequency and sequence uniformity on the Beauty dataset.}
    \label{fig:hyper}
\end{figure}

\subsection{Hyperparameter Experiment}\label{sec:4.4}

In this subsection, we explore the relationship between the performance of UniRec and two hyperparameters: the item frequency partition threshold and the user uniformity partition threshold. As shown in Figure \ref{fig:hyper}, we conduct experiments on the Amazon Beauty dataset, testing the impact of item frequency partition thresholds ranging from 40\% to 90\% (a), and sequence uniformity partition thresholds ranging from 30\% to 80\% (b). The results indicate that all tested partition thresholds yield good performance, but the most significant improvement occurs at specific values. For the Beauty dataset, the optimal split thresholds are 70\% for high-frequency items and 30\% for less-frequent items, while the ratio of uniform to non-uniform sequences is 60\% to 40\%. In summary, UniRec exhibits robust performance across different threshold settings, yet carefully selecting division thresholds can enhance the performance the most.

\begin{figure}
    \centering
    \includegraphics[width=0.95\columnwidth]{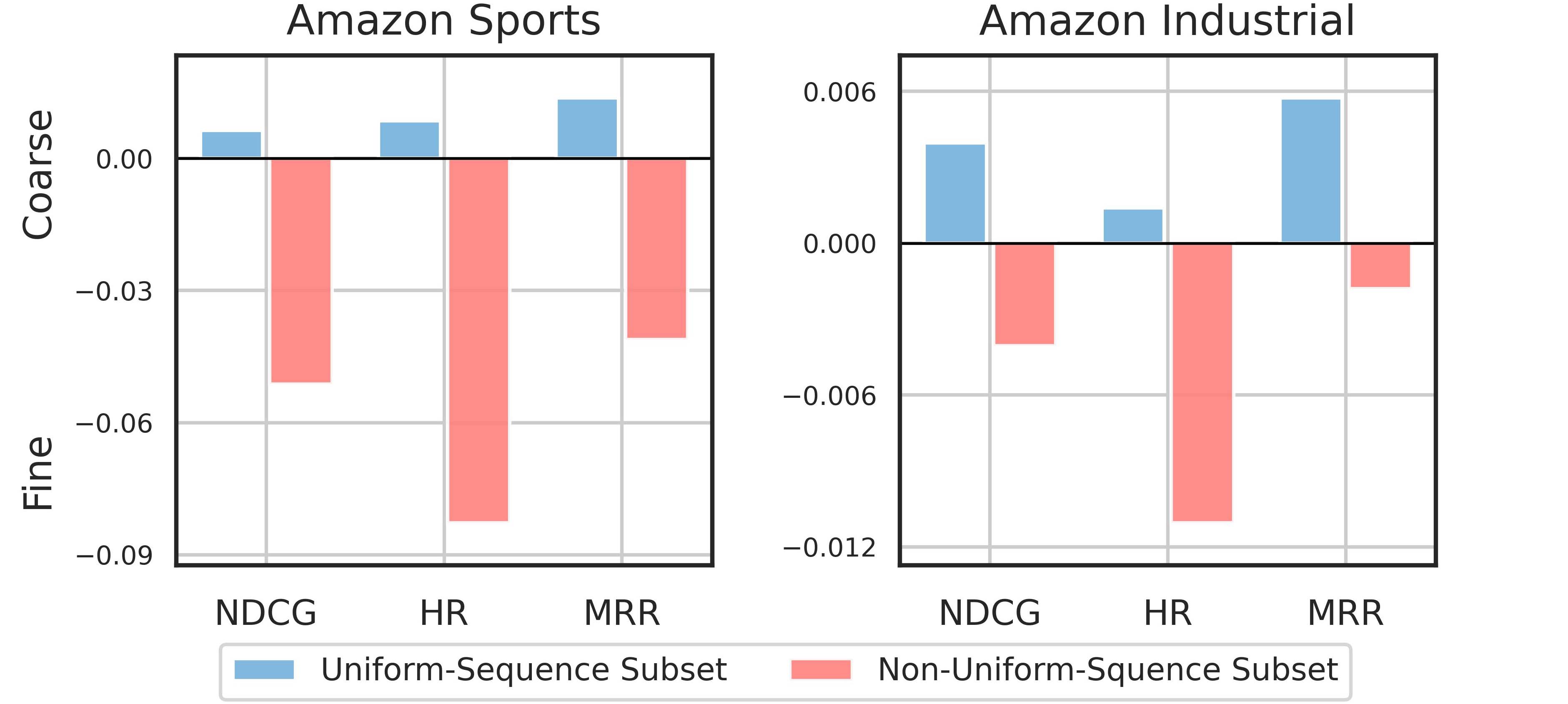}
    \caption{Time sensitivity comparison of uniform and non-uniform sequences on Amazon Sports and Amazon Industrial datasets.}
    \label{fig:test1}
\end{figure}

\subsection{Time Sensitivity Analysis}\label{sec:4.6}

\begin{figure}[ht]
    \centering
    \includegraphics[width=\columnwidth]{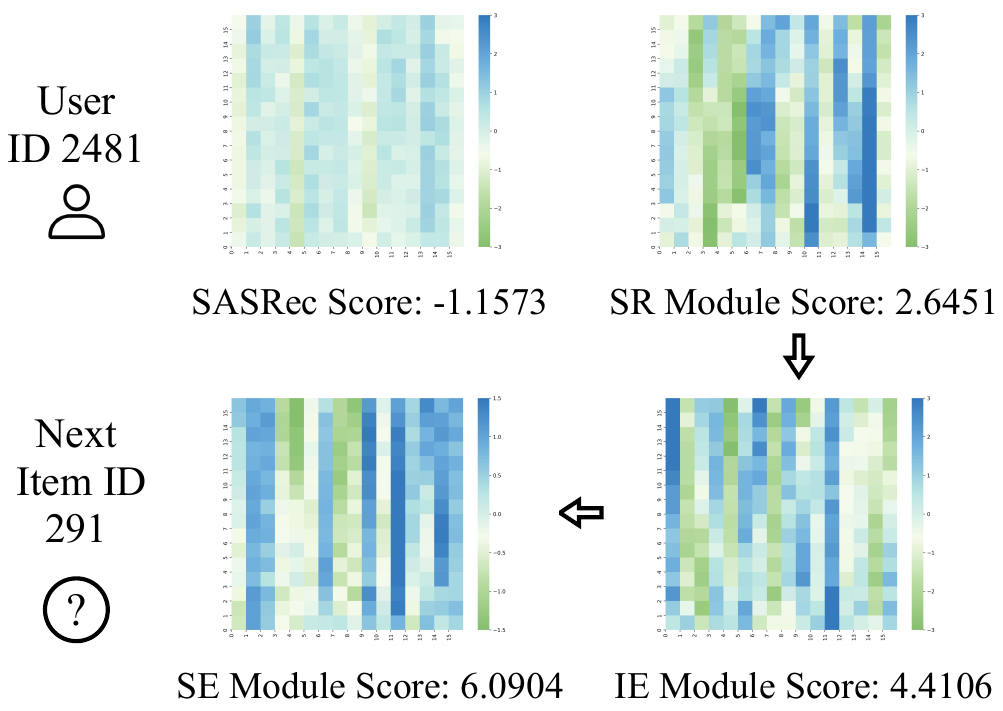}
    \caption{Prediction scores and corresponding sequence embedding heatmaps of a non-uniform sequence across different models and modules.}
    \label{fig:case}
\end{figure}

As mentioned in section \ref{sec:3}, we hypothesize that uniform sequences and non-uniform sequences may exhibit different dependencies on temporal information. In this subsection, to validate this hypothesis, we compare the effects of coarse-grained time modeling and fine-grained time modeling on both uniform and non-uniform sequence subsets. As shown in Figure \ref{fig:test1}, a positive score indicates that coarse-grained modeling outperforms fine-grained modeling, while the negative indicates the opposite. In both Amazon datasets, we observe that coarse-grained modeling performs better on uniform-sequence subsets, whereas fine-grained modeling is more effective on non-uniform-sequence subsets. For uniform sequences, user behavior patterns are more consistent, capturing global patterns can yield satisfactory predictive outcomes. Conversely, non-uniform sequences exhibit greater diversity and dynamism in user behavior, necessitating a fine-grained temporal encoding strategy to accurately model shifts and changes in user interests.

\subsection{Case Study} \label{sec:4.5}

We conduct a case study to illustrate the progressive enhancement of a non-uniform sequence through various models and modules. As shown in Figure \ref{fig:case}, we select a non-uniform sequence (user ID 2481) and demonstrate the changes in prediction scores for the next item (item ID 291) and the corresponding sequence embeddings after modeling with four different approaches: SASRec, SR module of UniRec, both the SR and IE modules, and the SR, IE, and SE modules. The progression of the model incorporating more modules is indicated by the arrows in the figure. SASRec shows a low prediction score, indicating its limited capability in handling sequences with significant interest drift. Adding the SR module significantly improves the model's predictive ability. The inclusion of the IE module brings further improvement, and the model achieves its best performance with the addition of the SE module. In the heatmaps, blue indicates larger positive values and green indicates smaller negative values. The transition in heatmap colors from SASRec to the enhanced models, with increasing contrast, demonstrates the model's growing ability to capture detailed information and features from various positions within the sequence.

\section{Related Works}
\subsection{Sequential Recommendation}
Sequential recommendation systems identify patterns in user behavior to predict future actions. Initially, Markov models \cite{rendle_factorizing_2010,johnson_enhancing_2017} are pivotal for analyzing transitions between states. The rise of deep learning leads to RNN models like GRU4Rec \cite{jannach_when_2017}, which improves predictions by capturing long-term dependencies \cite{hidasi_recurrent_2018,hidasi_session-based_2015,hidasi_parallel_2016}. Convolutional Neural Network (CNN \cite{lecun_deep_2015})-based models, such as Caser \cite{tang_personalized_2018}, improves recommendations by examining local behavior sequence patterns. Models like SHAN \cite{ying_sequential_2018} and STAMP \cite{liu_stamp_2018} effectively address shifts in user interests through memory strategies. Recently, attention mechanisms and Transformer-based models, like SASRec \cite{kang_self-attentive_2018} and Bert4Rec \cite{sun_bert4rec_2019}, have gained prominence. They leverage self-attention to understand complex sequence dependencies, while LightSANs \cite{fan_lighter_2021} introduces lightweight self-attention structures. The SSE-PT \cite{wu_sse-pt_2020} integrates personalized embeddings with Stochastic Shared Embeddings (SSE) \cite{wu_stochastic_2019}. Research also extends to cross-domain \cite{zhu2021cross,cao2022contrastive,guo2022reinforcement}, interpretable \cite{zhang2020explainable,huang2019explainable}, graph neural network \cite{hsu_retagnn_2021,ye_graph_2023,chang2021sequential,ma2020memory}, and contrastive learning approaches \cite{liu2021contrastive,yu2023self,yang2022knowledge,dang_uniform_2023} for sequential recommendations.

\subsection{Time-Aware Sequential Recommendation}
Time-aware systems incorporate timing to capture the dynamic nature of user preferences, offering more accurate and timely recommendations. These models surpass traditional ones by adapting recommendations to both the shifts in user preferences over time and their current interests \cite{ye_time_2020,fan2021continuous}. The TiSASRec \cite{li_time_2020} model innovatively adjusts self-attention weights based on the timing between actions, significantly improving performance. MEANTIME \cite{cho_meantime_2020} enriches time perception through diverse embedding techniques, whereas TASER \cite{ye_time_2020} explores both absolute and relative time patterns. TGSRec \cite{fan_continuous-time_2021} considers temporal dynamics in sequence patterns, and MOJITO \cite{tran_attention_2023} analyzes preferences from various temporal perspectives through a hybrid self-attention mechanism. FEARec \cite{du_frequency_2023} transitions sequence analysis from the time to the frequency domain, employing a hybrid attention mechanism and multitask learning for enhanced performance.

While these models ingeniously integrate temporal information, optimizing the use of such data remains a challenge. The diversity of data characteristics necessitates adaptable approaches for handling time intervals, timestamps, and cyclic patterns, given the varied and often irregular temporal behavior patterns among users. Recently, the TiCoSeRec \cite{dang_uniform_2023} introduces an innovative approach by considering sequence uniformity during the data augmentation phase, marking a deeper understanding of sequential recommendation data. While this model treats sequence uniformity as a target of data enhancement, it does not delve into modeling and analyzing this characteristic of the data further. In contrast, in this paper, we incorporate sequence uniformity into model construction. Our method not only addresses the limitations encountered by existing models when dealing with data of varied temporal distributions but also proposes a novel perspective for feature enhancement.
 
\section{Conclusion}

In this paper, we demonstrate that sequential recommendation algorithms perform better on uniform sequences and frequent items compared to non-uniform sequences and less-frequent items. To address this, we present a novel bidirectional enhancement architecture that leverages sequence uniformity and item frequency for feature enhancement, optimizing the performance of sequential recommendations. Additionally, we introduce a multidimensional time modeling method to better capture temporal information. Experimental results show that our method significantly outperforms twelve competitive models across four real-world datasets. To the best of our knowledge, this is the first work that utilizes the uniformity of sequences and frequency of items to enhance recommendation performance and it also indicates a promising direction and a new perspective for feature enhancement in future research.

\begin{acks}
This paper is sponsored by anonymous with Grant No.XXXXXXXX.
\end{acks}

\printbibliography

@inproceedings{kang_self-attentive_2018,
	
	title = {Self-{Attentive} {Sequential} {Recommendation}},
	
	
	
	
	booktitle = {2018 {IEEE} {International} {Conference} on {Data} {Mining} ({ICDM})},
	
	author = {Kang, Wang-Cheng and McAuley, Julian},
	
	year = {2018},
	pages = {197--206},
	
}

@inproceedings{sun_bert4rec_2019,
	
	title = {{BERT4Rec}: {Sequential} {Recommendation} with {Bidirectional} {Encoder} {Representations} from {Transformer}},
	
	shorttitle = {{BERT4Rec}},
	
	
	
	
	booktitle = {Proceedings of the 28th {ACM} {International} {Conference} on {Information} and {Knowledge} {Management}},
	
	author = {Sun, Fei and Liu, Jun and Wu, Jian and Pei, Changhua and Lin, Xiao and Ou, Wenwu and Jiang, Peng},
	
	year = {2019},
	pages = {1441--1450},
}

@inproceedings{wu_sse-pt_2020,
	
	title = {{SSE}-{PT}: {Sequential} {Recommendation} {Via} {Personalized} {Transformer}},
	
	shorttitle = {{SSE}-{PT}},
	
	
	
	
	booktitle = {Fourteenth {ACM} {Conference} on {Recommender} {Systems}},
	
	author = {Wu, Liwei and Li, Shuqing and Hsieh, Cho-Jui and Sharpnack, James},
	
	year = {2020},
	pages = {328--337},
	
}

@inproceedings{li_time_2020,
	
	title = {Time {Interval} {Aware} {Self}-{Attention} for {Sequential} {Recommendation}},
	
	
	
	
	
	booktitle = {Proceedings of the 13th {International} {Conference} on {Web} {Search} and {Data} {Mining}},
	
	author = {Li, Jiacheng and Wang, Yujie and McAuley, Julian},
	
	year = {2020},
	pages = {322--330},
	
}

@inproceedings{ye_time_2020,
	
	title = {Time {Matters}: {Sequential} {Recommendation} with {Complex} {Temporal} {Information}},
	
	shorttitle = {Time {Matters}},
	
	
	
	
	booktitle = {Proceedings of the 43rd {International} {ACM} {SIGIR} {Conference} on {Research} and {Development} in {Information} {Retrieval}},
	
	author = {Ye, Wenwen and Wang, Shuaiqiang and Chen, Xu and Wang, Xuepeng and Qin, Zheng and Yin, Dawei},
	
	year = {2020},
	pages = {1459--1468},
}

@inproceedings{cho_meantime_2020,
	
	title = {{MEANTIME}: {Mixture} of {Attention} {Mechanisms} with {Multi}-temporal {Embeddings} for {Sequential} {Recommendation}},
	
	shorttitle = {{MEANTIME}},
	
	
	
	
	booktitle = {Fourteenth {ACM} {Conference} on {Recommender} {Systems}},
	
	author = {Cho, Sung Min and Park, Eunhyeok and Yoo, Sungjoo},
	
	year = {2020},
	pages = {515--520},
	
}

@inproceedings{hidasi_recurrent_2018,
	
	title = {Recurrent {Neural} {Networks} with {Top}-k {Gains} for {Session}-based {Recommendations}},
	
	
	
	
	
	booktitle = {Proceedings of the 27th {ACM} {International} {Conference} on {Information} and {Knowledge} {Management}},
	
	author = {Hidasi, Balázs and Karatzoglou, Alexandros},
	
	year = {2018},
	pages = {843--852},
	
}

@article{hidasi_session-based_2015,
      title={Session-based Recommendations with Recurrent Neural Networks}, 
      author={Balázs Hidasi and Alexandros Karatzoglou and Linas Baltrunas and Domonkos Tikk},
      year={2016},
  journal={arXiv preprint arXiv:1511.06939},
}

@inproceedings{hidasi_parallel_2016,
	
	title = {Parallel {Recurrent} {Neural} {Network} {Architectures} for {Feature}-rich {Session}-based {Recommendations}},
	
	
	
	
	
	booktitle = {Proceedings of the 10th {ACM} {Conference} on {Recommender} {Systems}},
	
	author = {Hidasi, Balázs and Quadrana, Massimo and Karatzoglou, Alexandros and Tikk, Domonkos},
	
	year = {2016},
	pages = {241--248},
}

@inproceedings{jannach_when_2017,
	
	title = {When {Recurrent} {Neural} {Networks} meet the {Neighborhood} for {Session}-{Based} {Recommendation}},
	
	
	
	
	
	booktitle = {Proceedings of the {Eleventh} {ACM} {Conference} on {Recommender} {Systems}},
	
	author = {Jannach, Dietmar and Ludewig, Malte},
	
	year = {2017},
	pages = {306--310},
}

@inproceedings{wu_stochastic_2019,
	title = {Stochastic {Shared} {Embeddings}: {Data}-driven {Regularization} of {Embedding} {Layers}},
	volume = {32},
	
	booktitle = {Advances in {Neural} {Information} {Processing} {Systems}},
	
	author = {Wu, Liwei and Li, Shuqing and Hsieh, Cho-Jui and Sharpnack, James L},
	editor = {Wallach, H. and Larochelle, H. and Beygelzimer, A. and Alché-Buc, F. d' and Fox, E. and Garnett, R.},
	year = {2019},
}

@inproceedings{liu_stamp_2018,
	
	title = {{STAMP}: {Short}-{Term} {Attention}/{Memory} {Priority} {Model} for {Session}-based {Recommendation}},
	
	shorttitle = {{STAMP}},
	
	
	
	
	booktitle = {Proceedings of the 24th {ACM} {SIGKDD} {International} {Conference} on {Knowledge} {Discovery} \& {Data} {Mining}},
	
	author = {Liu, Qiao and Zeng, Yifu and Mokhosi, Refuoe and Zhang, Haibin},
	
	year = {2018},
	pages = {1831--1839},
}

@inproceedings{rendle_factorizing_2010,
	
	series = {{WWW} '10},
	title = {Factorizing personalized {Markov} chains for next-basket recommendation},
	
	booktitle = {Proceedings of the 19th international conference on {World} wide web},
	
	author = {Rendle, Steffen and Freudenthaler, Christoph and Schmidt-Thieme, Lars},
	
	year = {2010},
	pages = {811--820},
}

@inproceedings{johnson_enhancing_2017,
	
	series = {{WI} '17},
	title = {Enhancing long tail item recommendations using tripartite graphs and {Markov} process},
	
	booktitle = {Proceedings of the {International} {Conference} on {Web} {Intelligence}},
	
	author = {Johnson, Joseph and Ng, Yiu-Kai},
	
	year = {2017},
	pages = {761--768},
}

@article{lecun_deep_2015,
	title = {Deep learning},
	volume = {521},
	issn = {0028-0836, 1476-4687},
	
	
	
	number = {7553},
	
	journal = {Nature},
	author = {LeCun, Yann and Bengio, Yoshua and Hinton, Geoffrey},
	
	year = {2015},
	pages = {436--444},
	
}

@inproceedings{tang_personalized_2018,
	
	series = {{WSDM} '18},
	title = {Personalized {Top}-{N} {Sequential} {Recommendation} via {Convolutional} {Sequence} {Embedding}},
	
	booktitle = {Proceedings of the {Eleventh} {ACM} {International} {Conference} on {Web} {Search} and {Data} {Mining}},
	
	author = {Tang, Jiaxi and Wang, Ke},
	
	year = {2018},
	pages = {565--573},
	
}

@incollection{ying_sequential_2018,
	title = {Sequential recommender system based on hierarchical attention network},
	
	
	
	booktitle = {Sequential recommender system based on hierarchical attention network},
	collaborator = {Ying, H. and Zhuang, F. and Zhang, F. and Liu, Y. and Xu, G. and Xie, X. and Xiong, H. and Wu, J.},
	
	year = {2018},
	note = {ISSN: 1045-0823},
	pages = {3926--3932},
}

@inproceedings{fan_lighter_2021,
	
	series = {{SIGIR} '21},
	title = {Lighter and {Better}: {Low}-{Rank} {Decomposed} {Self}-{Attention} {Networks} for {Next}-{Item} {Recommendation}},
	
	shorttitle = {Lighter and {Better}},
	
	booktitle = {Proceedings of the 44th {International} {ACM} {SIGIR} {Conference} on {Research} and {Development} in {Information} {Retrieval}},
	
	author = {Fan, Xinyan and Liu, Zheng and Lian, Jianxun and Zhao, Wayne Xin and Xie, Xing and Wen, Ji-Rong},
	
	year = {2021},
	pages = {1733--1737},
}

@inproceedings{hsu_retagnn_2021,
	
	title = {{RetaGNN}: {Relational} {Temporal} {Attentive} {Graph} {Neural} {Networks} for {Holistic} {Sequential} {Recommendation}},
	
	shorttitle = {{RetaGNN}},
	
	
	
	
	booktitle = {Proceedings of the {Web} {Conference} 2021},
	
	author = {Hsu, Cheng and Li, Cheng-Te},
	
	year = {2021},
	pages = {2968--2979},
	
}

@inproceedings{ye_graph_2023,
	
	title = {Graph {Masked} {Autoencoder} for {Sequential} {Recommendation}},
	
	
	
	
	
	booktitle = {Proceedings of the 46th {International} {ACM} {SIGIR} {Conference} on {Research} and {Development} in {Information} {Retrieval}},
	
	author = {Ye, Yaowen and Xia, Lianghao and Huang, Chao},
	
	year = {2023},
	pages = {321--330},
	
}

@inproceedings{rahmani_incorporating_2023,
	
	title = {Incorporating {Time} in {Sequential} {Recommendation} {Models}},
	
	
	
	
	
	booktitle = {Proceedings of the 17th {ACM} {Conference} on {Recommender} {Systems}},
	
	author = {Rahmani, Mostafa and Caverlee, James and Wang, Fei},
	
	year = {2023},
	pages = {784--790},
	
}

@inproceedings{fan_continuous-time_2021,
	
	title = {Continuous-{Time} {Sequential} {Recommendation} with {Temporal} {Graph} {Collaborative} {Transformer}},
	
	
	
	
	
	booktitle = {Proceedings of the 30th {ACM} {International} {Conference} on {Information} \& {Knowledge} {Management}},
	
	author = {Fan, Ziwei and Liu, Zhiwei and Zhang, Jiawei and Xiong, Yun and Zheng, Lei and Yu, Philip S.},
	
	year = {2021},
	pages = {433--442},
	
}

@inproceedings{tran_attention_2023,
	
	title = {Attention {Mixtures} for {Time}-{Aware} {Sequential} {Recommendation}},
	
	
	
	
	
	booktitle = {Proceedings of the 46th {International} {ACM} {SIGIR} {Conference} on {Research} and {Development} in {Information} {Retrieval}},
	
	author = {Tran, Viet Anh and Salha-Galvan, Guillaume and Sguerra, Bruno and Hennequin, Romain},
	
	year = {2023},
	pages = {1821--1826},
	
}

@inproceedings{du_frequency_2023,
	
	title = {Frequency {Enhanced} {Hybrid} {Attention} {Network} for {Sequential} {Recommendation}},
	
	
	
	
	
	booktitle = {Proceedings of the 46th {International} {ACM} {SIGIR} {Conference} on {Research} and {Development} in {Information} {Retrieval}},
	
	author = {Du, Xinyu and Yuan, Huanhuan and Zhao, Pengpeng and Qu, Jianfeng and Zhuang, Fuzhen and Liu, Guanfeng and Liu, Yanchi and Sheng, Victor S.},
	
	year = {2023},
	pages = {78--88},
	
}

@article{dang_uniform_2023,
	title = {Uniform {Sequence} {Better}: {Time} {Interval} {Aware} {Data} {Augmentation} for {Sequential} {Recommendation}},
	volume = {37},
	copyright = {Copyright (c) 2023 Association for the Advancement of Artificial Intelligence},
	issn = {2374-3468},
	shorttitle = {Uniform {Sequence} {Better}},
	
	number = {4},
	
	journal = {Proceedings of the AAAI Conference on Artificial Intelligence},
	author = {Dang, Yizhou and Yang, Enneng and Guo, Guibing and Jiang, Linying and Wang, Xingwei and Xu, Xiaoxiao and Sun, Qinghui and Liu, Hong},
	
	year = {2023},
	note = {Number: 4},
	pages = {4225--4232},
	
}

@article{wang_survey_2022,
	title = {A {Survey} on {Session}-based {Recommender} {Systems}},
	volume = {54},
	issn = {0360-0300, 1557-7341},
	
	number = {7},
	
	journal = {ACM Computing Surveys},
	author = {Wang, Shoujin and Cao, Longbing and Wang, Yan and Sheng, Quan Z. and Orgun, Mehmet A. and Lian, Defu},
	
	year = {2022},
	pages = {1--38},
	
}

@inproceedings{huang2019explainable,
  title={Explainable interaction-driven user modeling over knowledge graph for sequential recommendation},
  author={Huang, Xiaowen and Fang, Quan and Qian, Shengsheng and Sang, Jitao and Li, Yan and Xu, Changsheng},
  booktitle={proceedings of the 27th ACM international conference on multimedia},
  pages={548--556},
  year={2019}
}

@article{zhang2020explainable,
  title={Explainable recommendation: A survey and new perspectives},
  author={Zhang, Yongfeng and Chen, Xu and others},
  journal={Foundations and Trends{\textregistered} in Information Retrieval},
  volume={14},
  number={1},
  pages={1--101},
  year={2020},
  publisher={Now Publishers, Inc.}
}

@article{guo2022reinforcement,
  title={Reinforcement learning-enhanced shared-account cross-domain sequential recommendation},
  author={Guo, Lei and Zhang, Jinyu and Chen, Tong and Wang, Xinhua and Yin, Hongzhi},
  journal={IEEE Transactions on Knowledge and Data Engineering},
  year={2022},
  publisher={IEEE}
}

@inproceedings{cao2022contrastive,
  title={Contrastive cross-domain sequential recommendation},
  author={Cao, Jiangxia and Cong, Xin and Sheng, Jiawei and Liu, Tingwen and Wang, Bin},
  booktitle={Proceedings of the 31st ACM International Conference on Information \& Knowledge Management},
  pages={138--147},
  year={2022}
}

@article{zhu2021cross,
  title={Cross-domain recommendation: challenges, progress, and prospects},
  author={Zhu, Feng and Wang, Yan and Chen, Chaochao and Zhou, Jun and Li, Longfei and Liu, Guanfeng},
  journal={arXiv preprint arXiv:2103.01696},
  year={2021}
}

@inproceedings{chang2021sequential,
  title={Sequential recommendation with graph neural networks},
  author={Chang, Jianxin and Gao, Chen and Zheng, Yu and Hui, Yiqun and Niu, Yanan and Song, Yang and Jin, Depeng and Li, Yong},
  booktitle={Proceedings of the 44th international ACM SIGIR conference on research and development in information retrieval},
  pages={378--387},
  year={2021}
}

@inproceedings{ma2020memory,
  title={Memory augmented graph neural networks for sequential recommendation},
  author={Ma, Chen and Ma, Liheng and Zhang, Yingxue and Sun, Jianing and Liu, Xue and Coates, Mark},
  booktitle={Proceedings of the AAAI conference on artificial intelligence},
  volume={34},
  number={04},
  pages={5045--5052},
  year={2020}
}

@article{liu2021contrastive,
  title={Contrastive self-supervised sequential recommendation with robust augmentation},
  author={Liu, Zhiwei and Chen, Yongjun and Li, Jia and Yu, Philip S and McAuley, Julian and Xiong, Caiming},
  journal={arXiv preprint arXiv:2108.06479},
  year={2021}
}

@article{yu2023self,
  title={Self-supervised learning for recommender systems: A survey},
  author={Yu, Junliang and Yin, Hongzhi and Xia, Xin and Chen, Tong and Li, Jundong and Huang, Zi},
  journal={IEEE Transactions on Knowledge and Data Engineering},
  year={2023},
  publisher={IEEE}
}

@inproceedings{yang2022knowledge,
  title={Knowledge graph contrastive learning for recommendation},
  author={Yang, Yuhao and Huang, Chao and Xia, Lianghao and Li, Chenliang},
  booktitle={Proceedings of the 45th international ACM SIGIR conference on research and development in information retrieval},
  pages={1434--1443},
  year={2022}
}

@inproceedings{fan2021continuous,
  title={Continuous-time sequential recommendation with temporal graph collaborative transformer},
  author={Fan, Ziwei and Liu, Zhiwei and Zhang, Jiawei and Xiong, Yun and Zheng, Lei and Yu, Philip S},
  booktitle={Proceedings of the 30th ACM international conference on information \& knowledge management},
  pages={433--442},
  year={2021}
}

@article{wang2019sequential,
  title={Sequential recommender systems: challenges, progress and prospects},
  author={Wang, Shoujin and Hu, Liang and Wang, Yan and Cao, Longbing and Sheng, Quan Z and Orgun, Mehmet},
  journal={arXiv preprint arXiv:2001.04830},
  year={2019}
}

@article{fang2020deep,
  title={Deep learning for sequential recommendation: Algorithms, influential factors, and evaluations},
  author={Fang, Hui and Zhang, Danning and Shu, Yiheng and Guo, Guibing},
  journal={ACM Transactions on Information Systems (TOIS)},
  volume={39},
  number={1},
  pages={1--42},
  year={2020},
  publisher={ACM New York, NY, USA}
}

@article{quadrana2018sequence,
  title={Sequence-aware recommender systems},
  author={Quadrana, Massimo and Cremonesi, Paolo and Jannach, Dietmar},
  journal={ACM computing surveys (CSUR)},
  volume={51},
  number={4},
  pages={1--36},
  year={2018},
  publisher={ACM New York, NY, USA}
}

@article{xu2019self,
  title={Self-attention with functional time representation learning},
  author={Xu, Da and Ruan, Chuanwei and Korpeoglu, Evren and Kumar, Sushant and Achan, Kannan},
  journal={Advances in neural information processing systems},
  volume={32},
  year={2019}
}

@misc{harper2015movielens,
  title={The MovieLens Datasets: History and Context},
  author={Harper, F. Maxwell and Konstan, Joseph A.},
  year={2015},
  eprint={1503.0863},
  archivePrefix={arXiv},
  primaryClass={cs.IR},
  note={arXiv:1503.0863},
}

@inproceedings{cho2011friendship,
  title={Friendship and mobility: user movement in location-based social networks},
  author={Cho, Eunjoon and Myers, Seth A. and Leskovec, Jure},
  booktitle={Proceedings of the 17th ACM SIGKDD international conference on Knowledge discovery and data mining},
  pages={1082--1090},
  year={2011},
  organization={ACM},
}

@inproceedings{mcauley2015image,
  title={Image-based recommendations on styles and substitutes},
  author={McAuley, Julian and Targett, Christopher and Shi, Qinfeng and van den Hengel, Anton},
  booktitle={Proceedings of the 38th International ACM SIGIR Conference on Research and Development in Information Retrieval},
  pages={43--52},
  year={2015},
  organization={ACM},
}

@inproceedings{he2016ups,
  title={Ups and downs: Modeling the visual evolution of fashion trends with one-class collaborative filtering},
  author={He, Ruining and McAuley, Julian},
  booktitle={Proceedings of the 25th International Conference on World Wide Web},
  pages={507--517},
  year={2016},
  organization={International World Wide Web Conferences Steering Committee},
}

@article{kingma2014adam,
  title={Adam: A Method for Stochastic Optimization},
  author={Kingma, Diederik P. and Ba, Jimmy},
  journal={arXiv preprint arXiv:1412.6980},
  year={2014},
}

@article{sarwar2001item,
  title={Item-based collaborative filtering recommendation algorithms},
  author={Sarwar, Badrul and Karypis, George and Konstan, Joseph and Riedl, John},
  journal={Proceedings of the 10th International Conference on World Wide Web},
  pages={285--295},
  year={2001},
  publisher={ACM}
}

@inproceedings{kim2019sequential,
  title={Sequential and Diverse Recommendation with Long Tail.},
  author={Kim, Yejin and Kim, Kwangseob and Park, Chanyoung and Yu, Hwanjo},
  booktitle={IJCAI},
  volume={19},
  pages={2740--2746},
  year={2019}
}

@inproceedings{liu2020long,
  title={Long-tail session-based recommendation},
  author={Liu, Siyi and Zheng, Yujia},
  booktitle={Proceedings of the 14th ACM Conference on Recommender Systems},
  pages={509--514},
  year={2020}
}


\end{document}